\documentclass[twocolumn,amssymb,nobibnotes,aps,prbstab, reprint,letterpaper,pre/printnumbers,amsmath,
superscriptaddress,
showpacs,floatfix]{revtex4-1}
\usepackage{graphicx,xcolor}
\usepackage{dcolumn}
\usepackage{bm}
\usepackage{graphics}
\usepackage{float}
\usepackage{epsfig}
\usepackage{multirow}
\usepackage{amsmath}
\usepackage{multirow}
\usepackage{tabularx}
\usepackage{mathrsfs}
\usepackage{tabularx}
\usepackage{tabulary}

\usepackage[pass,paperwidth=8.5in,paperheight=11in]{geometry}

\usepackage[linktocpage=true]{hyperref}
\usepackage{hyperref}
\usepackage{comment}
\usepackage{soul}
\usepackage{notes2bib}
\hypersetup{
  pdfnewwindow=true, 
  colorlinks=true,
  linkcolor=blue, 
  anchorcolor=blue,
  citecolor=blue, 
  filecolor=blue,
  menucolor=blue, 
  urlcolor=blue}

\usepackage[normalem]{ulem}
\usepackage{xcolor}
\usepackage{orcidlink}

\begin{document}
\newcommand\orangesout{\bgroup\markoverwith{\textcolor{orange}{\rule[0.5ex]{2pt}{0.4pt}}}\ULon}

\date{\today}

\title{First-principles investigation of elastic, vibrational, and thermodynamic properties of kagome metals CsM$_3$Te$_5$ (M = Ti, Zr, Hf) }

\author{Yifan Wei}
\affiliation{Department of Mechanical Engineering, University of Rochester, Rochester, New York 14627, USA}

\author{Arjyama Bordoloi\,\orcidlink{0009-0006-2760-3866}}
\affiliation{Department of Mechanical Engineering, University of Rochester, Rochester, New York 14627, USA}

\author{Chaon-En (Aaron) Chuang}
\affiliation{Department of Mechanical Engineering, University of Rochester, Rochester, New York 14627, USA}

\author{Sobhit Singh\,\orcidlink{0000-0002-5292-4235}}
\email{s.singh@rochester.edu}
\affiliation{Department of Mechanical Engineering, University of Rochester, Rochester, New York 14627, USA}
\affiliation{Materials Science Program, University of Rochester, Rochester, New York 14627, USA}
 
\begin{abstract}

Kagome metals are a unique class of quantum materials characterized by their distinct atomic lattice arrangement, featuring interlocking triangles and expansive hexagonal voids. These lattice structures impart exotic properties, including superconductivity, interaction-driven topological
many-body phenomena, and magnetism, among others. 
The kagome metal CsM$_3$Te$_5$ (where M = Ti, Zr, or Hf) exhibits both superconductivity and nontrivial topological electronic properties, offering a promising platform for exploring topological superconductivity. 
This study employs first-principles density functional theory calculations to systematically analyze the elastic, mechanical, vibrational, thermodynamic, and electronic properties of CsM$_3$Te$_5$ (M = Ti, Zr, Hf). 
Our calculations reveal that the studied compounds - CsTi$_3$Te$_5$, CsZr$_3$Te$_5$, and CsHf$_3$Te$_5$ - are ductile metals with elastic properties 
akin to the hexagonal Bi and Sb,
with average elastic constants, including a bulk modulus of 27\,GPa, a shear modulus of 11\,GPa, and Young's modulus of 29\,GPa. 
We observe peculiar dispersionless, flat, phonon branches in the vibrational spectra of these metals.
Additionally, we thoroughly analyze the symmetries of the zone-center phonon eigenvectors and predict vibrational fingerprints of the Raman- and infrared-active phonon modes. 
The analysis of thermodynamic properties reveals the Einstein temperature for CsTi\(_3\)Te\(_5\), CsZr\(_3\)Te\(_5\), and CsHf\(_3\)Te\(_5\) to be 66, 54, and 53\,K, respectively. 
Our orbital-decomposed electronic structure calculations reveal significant in-plane steric interactions and multiple Dirac band crossings near the Fermi level.
We further investigate the role of spin-orbit coupling effect on the studied properties. 
This theoretical investigation sheds light on the intriguing quantum behaviour of kagome metals.

\end{abstract}

\maketitle

\section{Introduction}
Kagome materials, distinguished by their unique lattice structure characterized by a distinctive hexagonal arrangement of interconnected triangles reminiscent of the traditional Japanese woven bamboo pattern known as ``kagome", exhibit a diverse array of rich electronic~\cite{Ye_nature, Mazin2014, GuoPRB2009,DennerPRL2021, Zhao2021, WulferdingPRR2022}, phononic~\cite{Yin2020, Korshunov2023}, topological~\cite{GuoPRB2009, Ghimire_Nature_2020, Mojarro_2024,Hu2022}, chiral~\cite{OhgushiPRB2000, Yu_PRB2012, Wang_PRB2021, Guo2022_Nature, KolincioPRL2023, Guo2024_npjQM}, and magnetic entangled properties~\cite{YanScience2011, Han2012, Nakatsuji2015, Kuroda2017, LiuScience2019, Kang2020, Xing2020, Legendre_PRR2020, Chen2024, Liu2018_NatPhy, Scheie2024}.~The symmetries present in these systems induces substantial magnetic and electronic frustration in triangularly coordinated magnetically active cations, leading to phenomena such as charge-density wave (CDW)~\cite{Luo_npj_quantum_2022,Ferrari2022, Arachchige2022, ZhouPRB2021, Liu_PRX2021, ChristensenPRB2021, Tan_PRL2023,Lee2024}, superconductivity~\cite{KoPRB2009, Ortiz2020, Ortiz2021, Zhao2021, Yin_nature_2022}, and chiral noncollinear magnetic orderings~\cite{Ghimire_Nature_2020, SinghPRB2023}. The study of kagome materials has gained significant attention following a seminal achievement by Ye \textit{et\,al.}~\cite{Ye_nature}, who reported  experimental realization of novel kagome materials. Their observation of exotic electronic behaviour and correlated topological phases has spurred a wave of research into kagome materials.

Among various kagome materials, the AV\(_3\)Sb\(_5\) family (A = K, Rb, Cs) has garnered significant attention due to their rich CDW states, giant anomalous Hall response, frustrated electronic structure, and associated superconductivity~\cite{Ortiz_PhysRevMaterials.3.094407,Subires_nature_com_2023,Uykur_npj_quantum_2022,Ritz_PhysRevB.107.205131,Ritz_PhysRevB_2.108.L100510,Ptok_PhysRevB.105.235134,SinghPRB2023}. Building on this interest, Si~\textit{et\,al.}~\cite{Si_PRB2022} have recently investigated a related family of kagome materials, CsM\(_3\)Te\(_5\) (M=Ti, Zr, Hf), which share a similar prototypic crystal structure with AV\(_3\)Sb\(_5\)~\cite{OrtizPRM2019}. 
Despite the absence of potential CDW transitions, 
this class of materials is unique due to the coexistence of superconductivity with nontrivial topological electronic properties. Notably, the superconducting transition temperatures of these materials are higher than those of the other members of the AV\(_3\)Sb\(_5\) family~\cite{Ortiz2020}.
The coexistence of superconductivity and topological electronic properties in these materials provides a unique platform to explore their interactions, potentially leading to the realization of novel quantum phases, such as Majorana zero-bound modes, with promising applications in fault-tolerant quantum computing~\cite{Kobayashi_PhysRevLett.115.187001, Tu_Materials_Today_Physics_2022100674, Qi_RevModPhys.83.1057}.
However, to advance the technological application of any material, a deep understanding of its elastic, mechanical, vibrational, and thermodynamic properties is crucial.

Despite extensive investigations on the superconductivity and nontrivial topological properties of CsM\(_3\)Te\(_5\)~\cite{Si_PRB2022}, studies on their elastic, mechanical, dynamical, and thermodynamic properties are still lacking. 
Hence, in this work, we focus on exploring these aspects of this relatively new class of kagome materials using first-principles density functional theory calculations.
Our calculations reveal that all three studied systems -- CsTi\(_3\)Te\(_5\), CsZr\(_3\)Te\(_5\), and CsHf\(_3\)Te\(_5\) -- are mechanically and dynamically stable, suggesting they may be synthesized experimentally. Additionally, they display ductile behavior with elastic moduli and elastic wave velocities comparable to those of hexagonal Bi and Sb.
We further predict the vibrational fingerprints of the studied phases to aid in their experimental identification in the future, and unveil some characteristic flat phonon branches in their phonon spectra. 
The analysis of thermodynamic properties reveals the Einstein temperature (\(\theta_E\)) for CsTi\(_3\)Te\(_5\), CsZr\(_3\)Te\(_5\), and CsHf\(_3\)Te\(_5\) to be 66, 54, and 53\,K, respectively. We also investigate the role of spin-orbit coupling on the studied properties.
All three studied systems exhibit metallic behavior with characteristic Dirac-like band crossings at the H and K points of the hexagonal Brillouin zone. Moreover, the in-plane \( p_x \) and \( p_y \) (d) orbitals of the Te (M) atom are higher in energy than the out-of-plane \( p_z \) (\( d_{xz}/d_{yz} \)) orbitals , indicating stronger in-plane steric interactions. This theoretical investigation sheds light on the intriguing electronic, elastic, mechanical, vibrational, and thermodynamic properties of these kagome metals.

\begin{figure}[!!t]
\centering
\includegraphics[width=1\columnwidth]{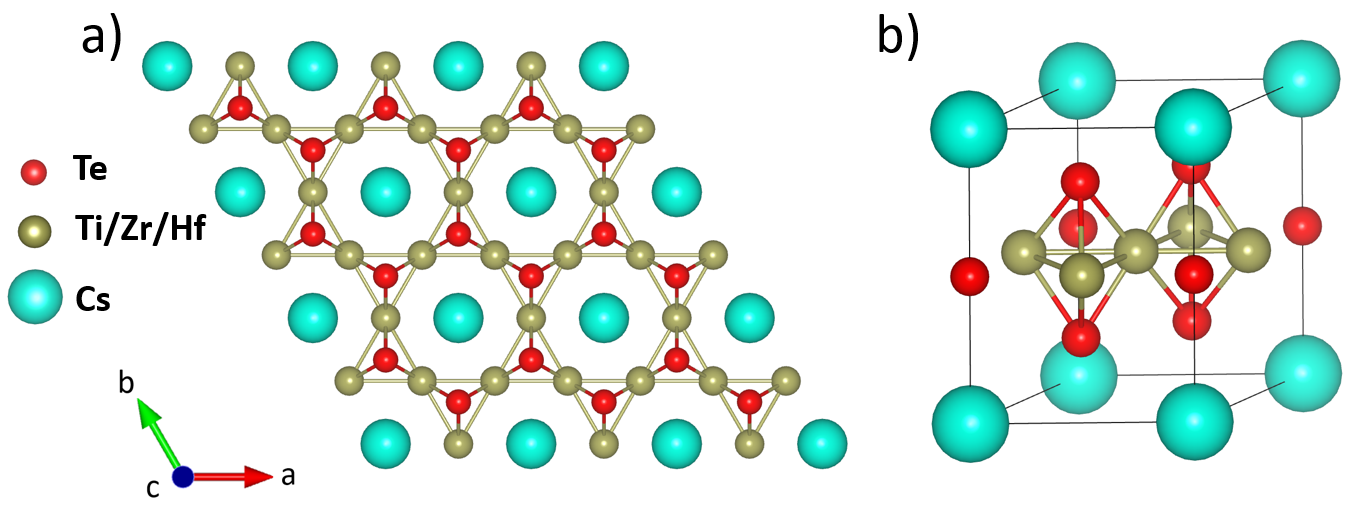}
\caption{Crystal structure of CsM\(_3\)Te\(_5\) ($P6/mmm$). a) Top view of the hexagonal crystal structure with cesium in the center and all other atoms arranged along the side and the vertex of the hexagon. b) Side view of the unit cell.}
\label{fig: crystal structure}
\end{figure}

\section{Computational Details}
First-principles density functional theory (DFT) calculations were conducted using the projector augmented-wave (PAW) method as implemented in the Vienna Ab initio Simulation Package (VASP)~\cite{Kresse96a, Kresse96b, KressePAW}. The exchange-correlation part of the Hamiltonian was treated using generalized-gradient approximation as parametrized by Perdew, Burke, and Ernzerhof for solids (PBEsol)~\cite{PBEsol}, with a kinetic energy cutoff for the plane wave basis set as 600\,eV. The energy convergence criterion for the electronic self-consistent calculations was set to ${10}^{-7}$\,eV. A complete relaxation of the lattice parameters as well as inner-atomic coordinates was performed until an optimized ground state was achieved, with residual Hellmann-Feynman forces less than ${10}^{-3}$\,eV/\AA~per atom. CsM$_3$Te$_5$(M=Ti, Zr, Hf)) crystallizes into $P6/mmm$ (no.\,191) symmetry with the optimized lattice parameters listed in Table \ref{tab:lattice parameters}. 
Figure \ref{fig: crystal structure} represents the crystal structure of CsM$_3$Te$_5$ generated using Vesta~\cite{vesta}. All calculations, except for phonon calculations, included spin-orbit coupling (SOC) effects. A $\Gamma$-centered k-mesh of size 8\,$\times$\,8\,$\times$\,6 was employed to sample the Brillouin zone. In the PAW pseudopotential, contributions from nine electrons for Cs (5s$^2$5p$^6$6s$^1$) and six electrons for Te (5s$^2$5p$^4$) were considered. For the transition metal atoms, twelve electrons were considered for each of Ti (3s$^2$3p$^6$3d$^2$4s$^2$), Zr (4s$^2$4p$^6$4d$^2$5s$^2$), and Hf (5s$^2$5p$^6$5d$^2$6s$^2$).

The dynamical stability of the studied systems was investigated through phonon calculations in 2\,$\times$\,2\,$\times$\,1 supercells using the finite displacement method as implemented in VASP. Post-processing of phonon data was performed using {\sc PHONOPY} software~\cite{phonopy}. 
The elastic constants C$_{ij}$ were converged to within 1\,GPa by increasing the k-mesh size.
A detailed analysis of the elastic properties was conducted using the {\sc MechElastic} Python package~\cite{MechElastic} including the calculation of various physical quantities, such as elastic wave velocities and Debye temperature. 
Given the significant role of SOC in the electronic and vibrational behaviors of the CsM$_3$Te$_5$ system, a detailed investigation was carried out to understand its impact on the electronic, elastic, and mechanical properties. {\sc PyProcar} software~\cite{pyprocar} was used for post-processing of electronic structure data.

\begin{table}[tb]
\centering
\caption{DFT-optimized (PBEsol) lattice parameters of CsM$_3$Te$_5$(M=Ti, Zr, Hf)}
\label{tab:lattice parameters}
\begin{tabular}{| c | c | c | c |}
\hline
\textbf{} & ~\textbf{CsTi$_3$Te$_5$} ~& ~\textbf{CsZr$_3$Te$_5$}~ & ~\textbf{CsHf$_3$Te$_5$}~ \\
\hline\
a = b (\AA)~~ & 5.973 & 6.280 & 6.229\\
c (\AA)~~ & 8.372 & 8.493 & 8.443 \\
\hline
\end{tabular}
\end{table}

\section{Results and Discussions}

\subsection{Elastic and Mechanical Properties}

The macroscopically measurable quantities of materials, such as Young's modulus (E), Poisson's ratio ($\nu$), bulk modulus (K), and shear modulus (G), are key indicators of their elastic and mechanical properties. Understanding these properties is essential for assessing the potential technological applications of materials.
While analyzing the elastic and mechanical properties of materials, it is essential to first discuss the elastic stiffness constants \(C_{ijkl}\) and define their relationship with the macroscopically measurable quantities. This can be obtained using the generalized stress-strain Hooke's law~\cite{Hooke}, which can be expressed as,
\begin{equation}
\sigma_{ij} = C_{ijkl} \epsilon_{kl},
\end{equation}
where ${\sigma}_{ij}$ and $\epsilon_{kl}$ represent homogeneous two-rank stress and strain tensors, respectively, and \(C_{ijkl}\) represents the fourth-rank elastic stiffness tensor. Utilizing the symmetry of the stress ${\sigma}_{ij}$  
and strain $\epsilon_{kl}$ tensors, we can reduce the total number of independent \(C_{ijkl}\)  coefficients from 81 to 36.
Further, using the crystal symmetry operations present in hexagonal structures, the total number of elastic stiffness constants for the investigated systems can be reduced from 36 to 5 within the Voigt notation~\cite{Hooke}.

\begin{table}[ht]
\renewcommand*{\arraystretch}{1.25}
\caption{Elastic constants (C$_{ij}$) computed with and without SOC, {\it i.e.}, PBE+SOC and PBE, respectively. Values of C$_{ij}$ (in GPa) obtained without PBE+SOC are enclosed in parentheses.}
\label{tab:elastic constants}
\begin{tabularx}{0.9\columnwidth}{c|c|c|c|c|c}
\hline
Composition~~~& ~~C$_{11}$ ~~&~~ C$_{12}$~~ & ~~C$_{13}$~~ & ~~C$_{33}$~~ & ~~C$_{44}$ ~~\\
\hline\
CsTi$_3$Te$_5$ & 78.8 & 17.3 & 10.5 & 32.5 & 5.0\\
& (79.4) & (16.9) & (10.6) & (32.7) & (5.4)\\
CsZr$_3$Te$_5$ & 72.0 & 19.1 & 10.8 & 34.3 & 0.2\\
 & (71.8) & (19.4) & (10.6) & (34.0) & (0.5)\\
CsHf$_3$Te$_5$ & 75.1 & 22.5 & 11.2 & 34.8 & 1.1\\
 & (75.2) & (23.5) & (10.8) & (35.0) & (1.2)\\
\hline
\end{tabularx}
\end{table}

Table \ref{tab:elastic constants} presents the computed values of the relevant elastic constants for CsM$_3$Te$_5$ (M = Ti, Zr, Hf), both with and without including SOC effects in our calculations. Strikingly, the inclusion of SOC does not significantly affect the values of the elastic constants, as evident from Table \ref{tab:elastic constants}. However, the C$_{ij}$  values computed with SOC are considered for the rest of the elastic-properties-dependent calculations to ensure better accuracy.

\begin{figure}[!!b]
\centering
\includegraphics[width=1\columnwidth]{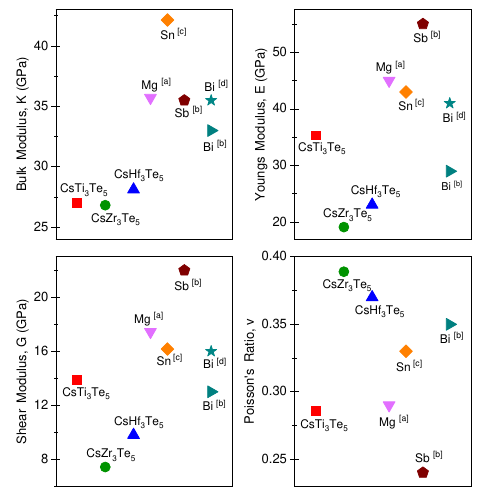}
\caption{Elastic constant of CsM$_3$Te$_5$ compounds with SOC and comparison among other materials.\\
\textsuperscript{a} Reference \cite{ASM} experiment data of Mg\newline{}
\textsuperscript{b} Reference \cite{Singh_elastic} data of Sb and Bi from LDA+SOC calculations\\
\textsuperscript{c} Reference \cite{fundamentals_of_Mat-sci} data of Sn from Fundamentals of Materials Science and Engineering: An Integrated Approach\\
\textsuperscript{d} Reference \cite{exp_data_4.2K_Bi} experiment data of Bi's elastic properties at 4.2K}
\label{fig:elastic_property}
\end{figure}

Next, we studied the mechanical stability of the three investigated systems using the Born-Huang mechanical stability criteria for hexagonal structures~\cite{MechElastic}. According to these criteria, a crystal is mechanically stable if it possesses the lowest Gibbs free energy in its relaxed state, {\it i.e.}, the state with no external loads, as compared to any other deformed state that can be reached by applying an infinitesimally small strain. Consequently, the crystal's elastic stiffness matrix, \(C_{ij}\), must be positive definite, meaning all the eigenvalues of \(C_{ij}\) must be positive, and the matrix must be symmetric. 
Furthermore, 
$C_{11} - C_{12} > 0$, $2 C_{13}^2 < C_{33}(C_{11} + C_{12}),$ and $C_{44} > 0$ are required conditions for mechanical stability~\cite{MechElastic}.
As can be inferred from Table \ref{tab:elastic constants}, all three studied systems meet the above-mentioned conditions for mechanical stability.


\begin{table*}[!!t]
\renewcommand*{\arraystretch}{1.25}
\caption{Elastic constants of CsM$_3$Te$_5$ calculated with inclusion of SOC effects} 
\begin{tabularx}{1.99\columnwidth}{c|c|c|c|c|c|c|c|c|c}
\hline
Composition ~~ & ~K (GPa)~ & ~G (GPa) ~&~E (GPa) ~& ~~ $\nu$ ~~&~~ $v_l$ (m/s) ~~& ~~$v_t$ (m/s)~~&~~ $v_m$ (m/s)~~&~$\theta_{\text{Debye}}$ (K) ~& ~$\theta_E = 6\,T_0$ (K)\\
\hline
CsTi$_3$Te$_5$ & 27.2 & 14.2 & 36.1 & 0.28 & 2803.1 & 1666.6 & 1733.7 & 168.5 & 66.1\\
CsZr$_3$Te$_5$ & 26.8 & 8.1 & 21.0 & 0.38 & 2506.6 & 1163.0 & 1309.9 & 122.5 & 54.7\\
CsHf$_3$Te$_5$ & 27.0 & 11.2 & 28.6 & 0.33 & 2298.8 & 1091.8 & 1228.2 & 115.8 & 53.1\\
\hline
\end{tabularx}

\label{tab:data}
\end{table*}

Once the \(C_{ij}\) constants are computed, all four elastic moduli (K, E, G, $\nu$) can be determined using their relationships with the elastic constants. First, the Voigt-Reuss-Hill (VRH)~\cite{Hill_Proc_Phy_Soc_A_1952} averaging scheme is used to calculate \(K\) and \(G\). Then, \(\nu\) and \(E\) are determined using their standard relationships with \(K\), \(G\), and the elastic constants, as implemented within the {\sc MechElastic } package~\cite{MechElastic}. 
The computed values of the mechanical and elastic moduli of the studied kagome metals are listed in Table~\ref{tab:data}. These values are crucial for further study and practical applications of these compounds.

\begin{figure}[!!t]
\centering
\includegraphics[width=1\columnwidth]{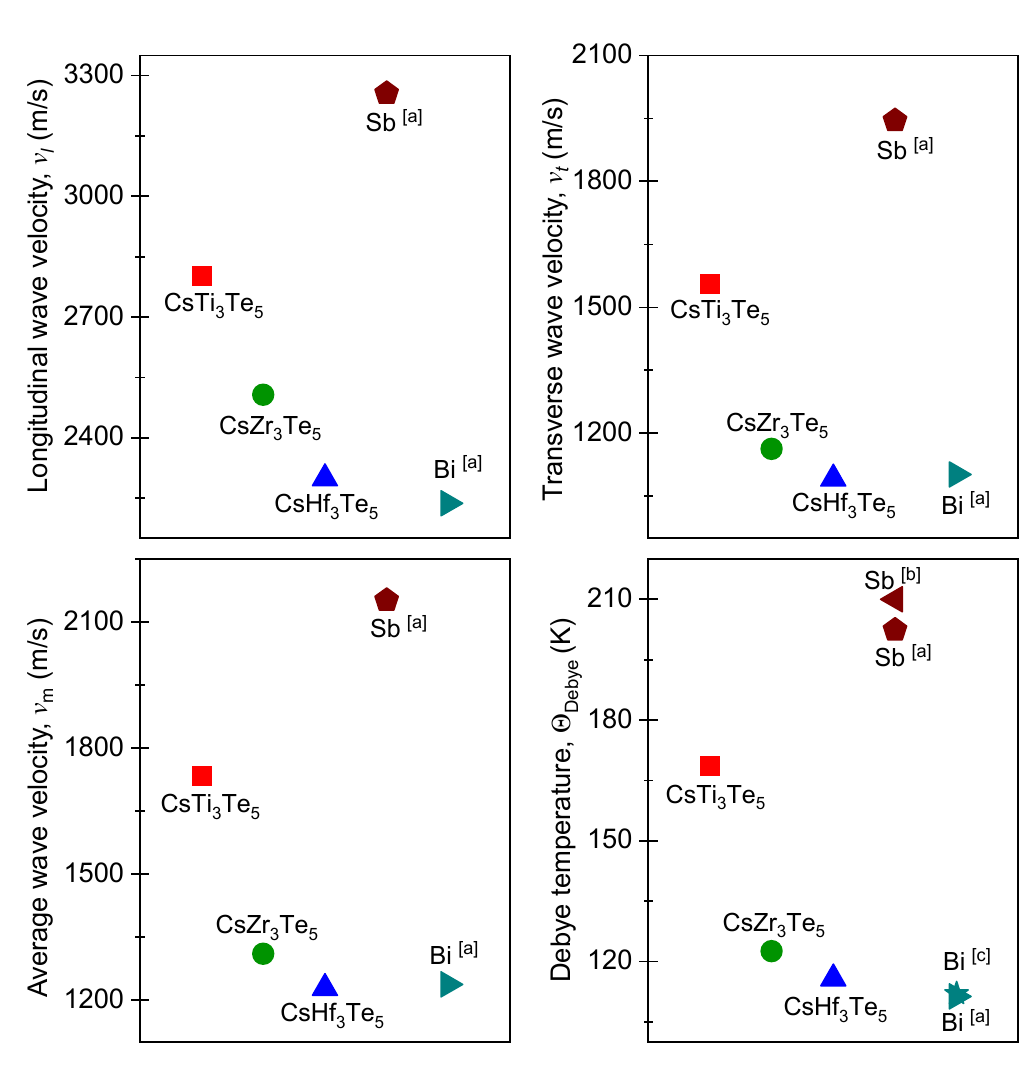}
\caption{Wave velocities and Debye temperatures of CsM$_3$Te$_5$ compounds with SOC and comparison among other materials.\\ 
\textsuperscript{a} Reference \cite{Singh_elastic} data of Sb and Bi from LDA+SOC calculations\\
\textsuperscript{b} Reference \cite{exp_data_Sb_wave} experiment data of Sb's Debye Temperature
\textsuperscript{c} Reference \cite{exp_data_Bi_wave} experiment data of Bi's Debye Temperature}
\label{fig:elastic_vel}
\end{figure}

Figure~\ref{fig:elastic_property} illustrates the values of the elastic moduli for the three investigated systems along with those of some commonly studied materials, providing a comparative analysis. As shown in Fig.~\ref{fig:elastic_property}, the bulk modulus does not exhibit significant variation with changing the M atom in CsM$_3$Te$_5$. The elastic moduli of these systems are comparable to those of Bi and Sb, which also have hexagonal structures. A comparison is also made with \(\beta\)-Sn, which is a widely used engineering material. Furthermore, the ductility or brittleness of a material can be assessed by calculating the ratio of \(K\) to \(G\). A value above 1.7 indicates a ductile behaviour~\cite{MechElastic}. All three investigated kagome metals are ductile, as expected for metals.

Additionally, we compute the elastic wave velocities and Debye temperature, which are crucial parameters for technological applications of kagome metals. The values of longitudinal (\(v_l\)), transverse (\(v_t\)), and average (\(v_m\)) elastic wave velocities are determined using the MechElastic package~\cite{MechElastic}, employing the following formulae:
\begin{equation}
    v_l = \sqrt{\frac{3K+4G}{3\rho}},
\end{equation}
\begin{equation}
    v_t = \sqrt{\frac{G}{\rho}}, ~~\text{and}
\end{equation}
\begin{equation}
    v_m = \Big[\frac{1}{3}(\frac{2}{v_t^3}+\frac{1}{v_l^3}\Big]^{-1/3}.
\end{equation}
Here, $\rho$ is the density of material. 
Further, the Debye temperature (\(\theta_{\text{Debye}}\)) of the studied systems is determined from the calculated elastic wave velocities and density $\rho$ (Eq.~\ref{eq:debye}), and it correlates with several important physical properties such as elastic constants, specific heat, ultrasonic wave velocities, and melting temperatures~\cite{MechElastic}. Since the acoustic phonons are the only contributors to the specific heat at low temperatures, hence the value of \(\theta_{\text{Debye}}\) obtained from specific heat measurements at low temperatures are expected to be consistent with that derived from the elastic constants. The \(\theta_{\text{Debye}}\) formula reads~\cite{ANDERSON1963909}:
\begin{equation}
    \theta_{Debye} = \frac{h}{k_B}\Big[\frac{3q}{4\pi}\frac{N\rho}{M}\Big]^{1/3}v_m, 
\label{eq:debye}
\end{equation}

where \( h \) is Planck's constant, \( k_B \) is Boltzmann's constant, \( q \) is the total number of atoms in the formula unit, \( N \) is Avogadro's number,  and \( M \) is the molecular weight of the solid.

Figure \ref{fig:elastic_vel} illustrates the calculated values of \( v_l \), \( v_t \), \( v_m \), and \( \theta_{\text{Debye}} \) (also, see Table~\ref{tab:data}). Additionally, comparisons are drawn with standard hexagonal systems including Sb and Bi. From Figure \ref{fig:elastic_vel}, it is evident that CsTi$_3$Te$_5$ exhibits the highest values among the three kagome materials for \( v_l \), \( v_t \), \( v_m \), and \( \theta_{\text{Debye}} \). This is expected because as the atomic number increases from Ti to Zr to Hf, the atomic mass increases, lowering the phonon frequencies as discussed below.

\subsection{Lattice Dynamics}

In order to understand the lattice dynamics and identify the vibrational spectroscopic signatures of the studied kagome metals, we perform phonon calculations within the finite-displacement method. DFT (PBEsol) calculated phonon spectra, along with the atom-resolved phonon density of states (PDOS) are shown in Figure~\ref{fig:phonon_spectra}. We observe all stable phonon frequencies in the whole Brillouin zone. 
With increasing atomic mass of $M$ atom (Ti $\rightarrow$ Zr $\rightarrow$ Hf), we note a systematic decrease in the phonon frequencies due to their $1/\sqrt{\text{mass}}$ dependence. 

\begin{figure}[!!t]
\centering
\includegraphics[width=9 cm]{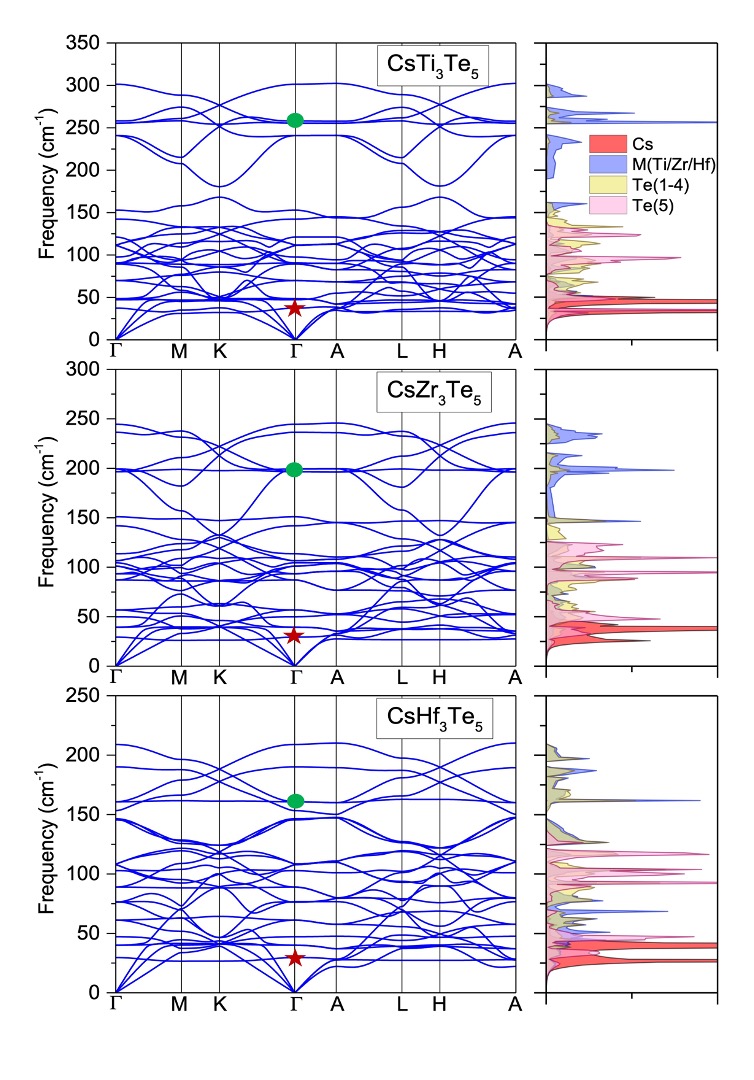}
\caption{DFT (PBEsol) calculated phonon spectrum plotted along with the atom-resolved phonon density of states (PDOS) for CsTi$_3$Te$_5$, CsZr$_3$Te$_5$, and CsHf$_3$Te$_5$.}
\label{fig:phonon_spectra}
\end{figure}

\begin{figure}[!!b]
\centering
\includegraphics[width=7.5 cm]{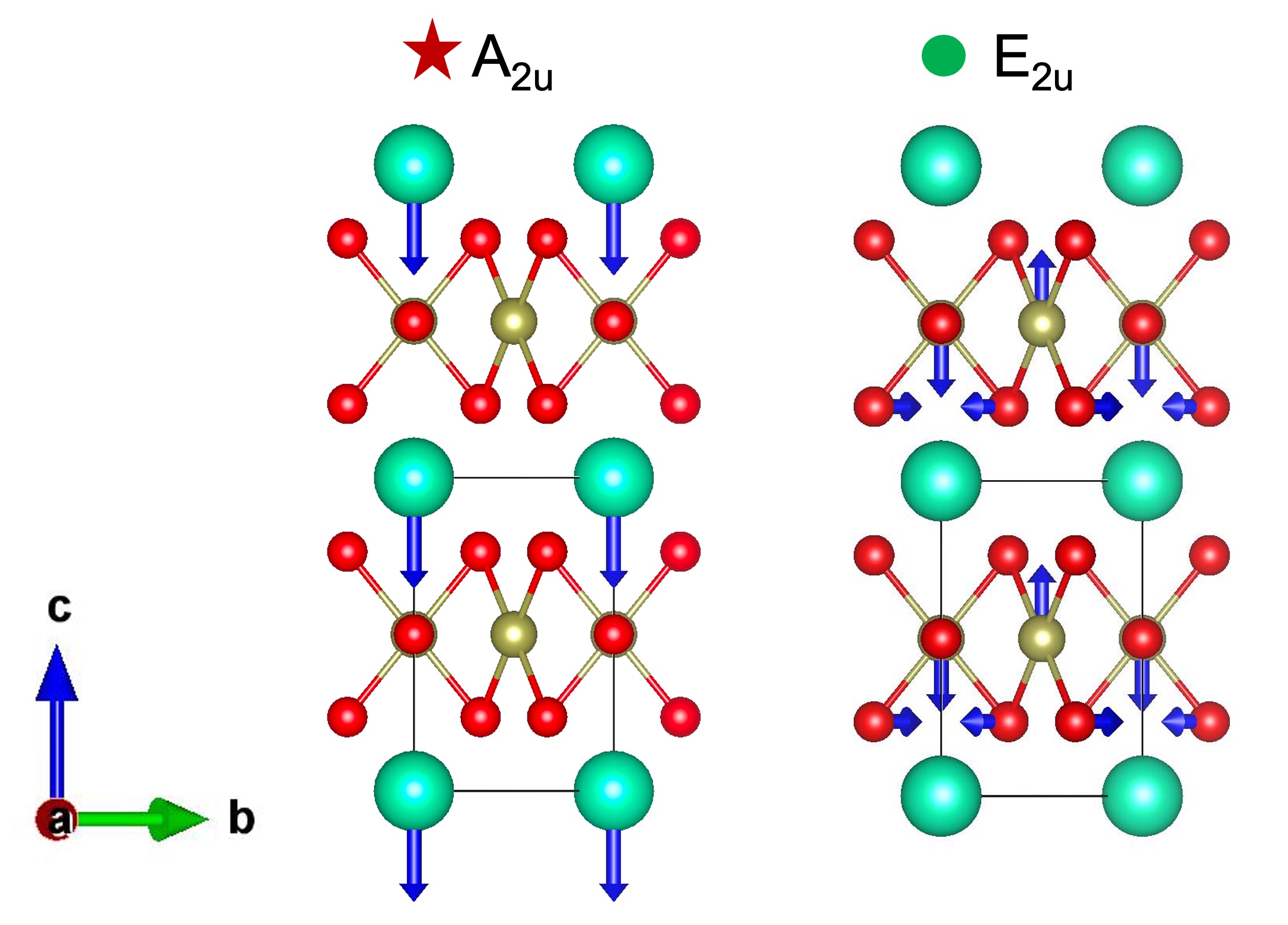}
\caption{Atomic displacement patterns corresponding to the Einstein modes $\emph{A}\textsubscript{2u}$~(left) and $\emph{E}\textsubscript{2u}$~(right) present in CsM$_3$Te$_5$. Color coding of atoms is the same as in Figure~\ref{fig: crystal structure}.}
\label{fig:atomic_dispalcement}
\end{figure}

\begin{figure}[ht!]
\centering
\includegraphics[width=1\columnwidth]{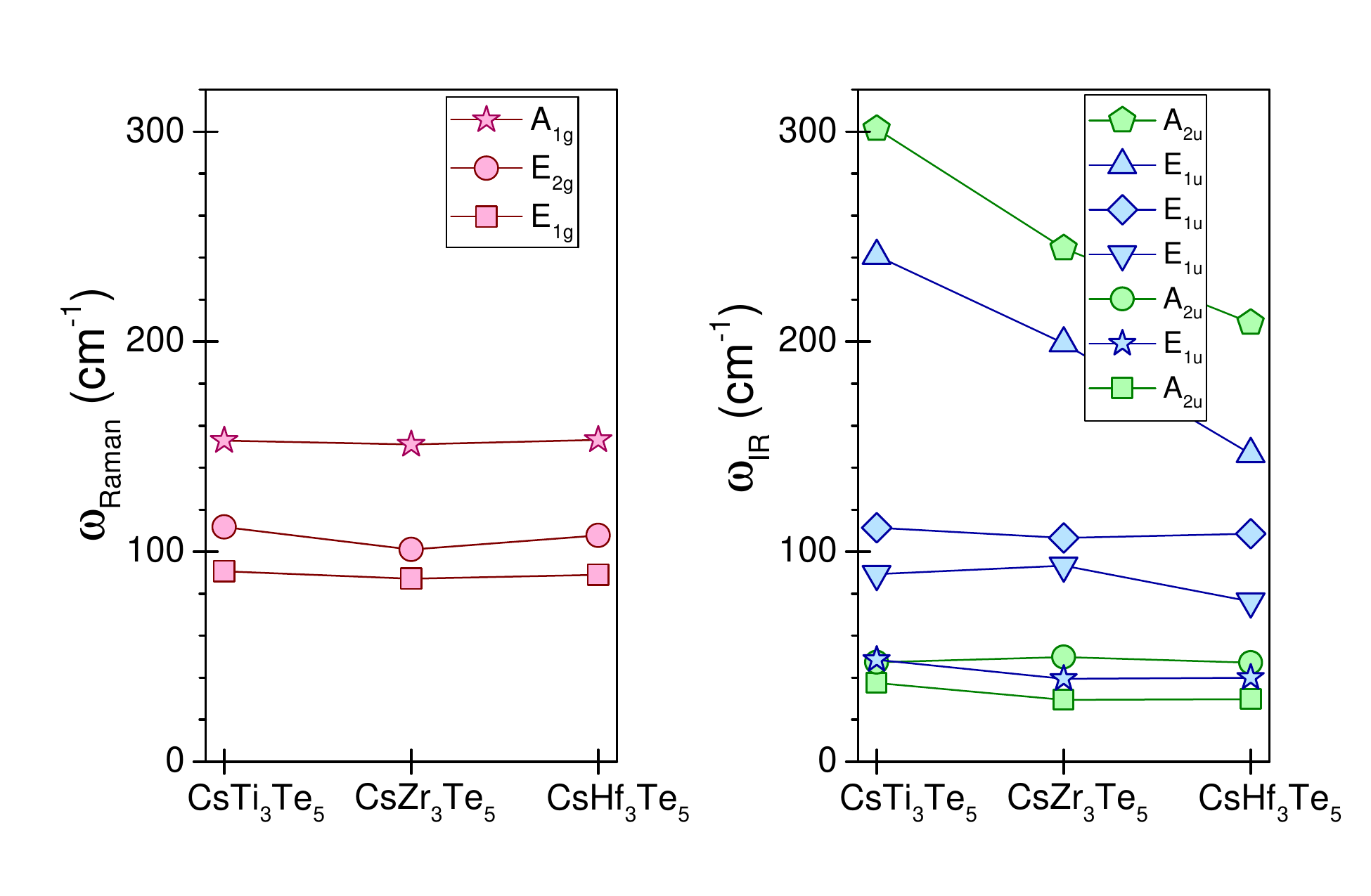}
\caption{Raman- (left) and IR-active (right) phonon frequencies calculated for CsTi$_3$Te$_5$, CsZr$_3$Te$_5$, and CsHf$_3$Te$_5$.}
\label{fig: raman_ir}
\end{figure}

Interestingly, we notice several dispersionless, ``almost flat", phonon branches extending throughout the whole Brillouin zone. These branches yield a sharp spike in the calculated PDOS at certain frequencies, as shown in Fig.~\ref{fig:phonon_spectra}. 
Such flat phonon branches are characteristics of the Einstein's oscillators. Two particularly noteworthy Einstein's oscillators are marked in Fig.~\ref{fig:phonon_spectra} using red and green symbols. 
One, the lowest frequency optical phonon branch ($\emph{A}\textsubscript{2u}$ symmetry at $\Gamma$) near 25\,cm$^{-1}$, highlighted using red star , corresponds to the out-of-plane vibrations of the Cs atoms in the lattice. 
Another interesting Einstein-like phonon branch is the highest frequency optical mode ($\emph{E}\textsubscript{2u}$ symmetry at $\Gamma$) highlighted using green circle. This mode corresponds to a complex out-of-plane as well in-plane vibration of the $M$ and Te atoms.
The atomic displacement patterns for  both these Einstein modes are shown in Fig.~\ref{fig:atomic_dispalcement}.

In the calculated PDOS, we separate the atomic contributions from the two groups of symmetry nonequivalent Te atoms, {\it i.e.}, one those who lie within the kagome plane (Te5) and other which do not lie in the kagome plane (Te1, Te2, Te3, Te4), as shown in Fig.~\ref{fig: crystal structure}. 
We observe that out-of-plane Te atoms are relatively strongly coupled to the vibrations of the M atoms, whereas in-plane Te atoms couple moderately with the vibrations of the M and/or Cs atoms. 

To identify the vibrational spectroscopic signatures of the studied compounds, we examine the zone-center phonon eigenvectors. There are nine atoms per unit cell in CsM$_3$Te$_5$ ($P6/mmm$) resulting in a total of 27 phonon modes. All the allowed phonon eigen modes at the Brillouin-zone center can be described using the following irreducible representations. 
\begin{equation}\label{eq1}
\begin{aligned}
\Gamma\textsubscript{vib} = \emph{A}\textsubscript{1g} \oplus
4\emph{A}\textsubscript{2u} \oplus  \emph{B}\textsubscript{1g} \oplus 
\emph{B}\textsubscript{1u} \oplus 2\emph{B}\textsubscript{2u}  \oplus
 2\emph{E}\textsubscript{2u} \\ \oplus \, \emph{E}\textsubscript{2g} \oplus 5\emph{E}\textsubscript{1u} \oplus 
 \emph{E}\textsubscript{1g}. 
\end{aligned}
\end{equation}

Among these 27 phonon modes, three are acoustic, denoted as 
$\Gamma\textsubscript{acoustic}$ = $\emph{A}\textsubscript{2u} \oplus  \emph{E}\textsubscript{1u}$.
The remaining 24 modes are optical, represented as 
$\Gamma\textsubscript{optic}$ = 
$ \emph{A}\textsubscript{1g} \oplus 
3\emph{A}\textsubscript{2u} \oplus 
\emph{B}\textsubscript{1g}  \oplus 
\emph{B}\textsubscript{1u}  \oplus 
2\emph{B}\textsubscript{2u}  \oplus 
2\emph{E}\textsubscript{2u}  \oplus 
\emph{E}\textsubscript{2g}  \oplus 
4\emph{E}\textsubscript{1u}  \oplus 
\emph{E}\textsubscript{1g} $. 
Specifically, the 
 $\emph{A}\textsubscript{1g}$, 
 $\emph{E}\textsubscript{2g}$, and $\emph{E}\textsubscript{1g}$ modes are Raman active, while the  
 $\emph{A}\textsubscript{2u}$ and $\emph{E}\textsubscript{1u}$ modes are infrared (IR) active. The rest of the modes are silent.
Figure~\ref{fig: raman_ir} illustrates the calculated Raman- and IR-active phonon frequencies for CsTi$_3$Te$_5$, CsZr$_3$Te$_5$, and CsHf$_3$Te$_5$. 
As expected, the  $\emph{A}\textsubscript{2u}$ and $\emph{E}\textsubscript{1u}$ phonon modes associated with the vibrations of the M atom (where M is Ti, Zr, or Hf) progressively soften as the atomic mass of M increases. In contrast, the frequencies of the other phonon modes remain largely unchanged.

\subsection{Specific Heat}

\begin{figure}[!!t]
\centering
\includegraphics[width=9.5 cm]{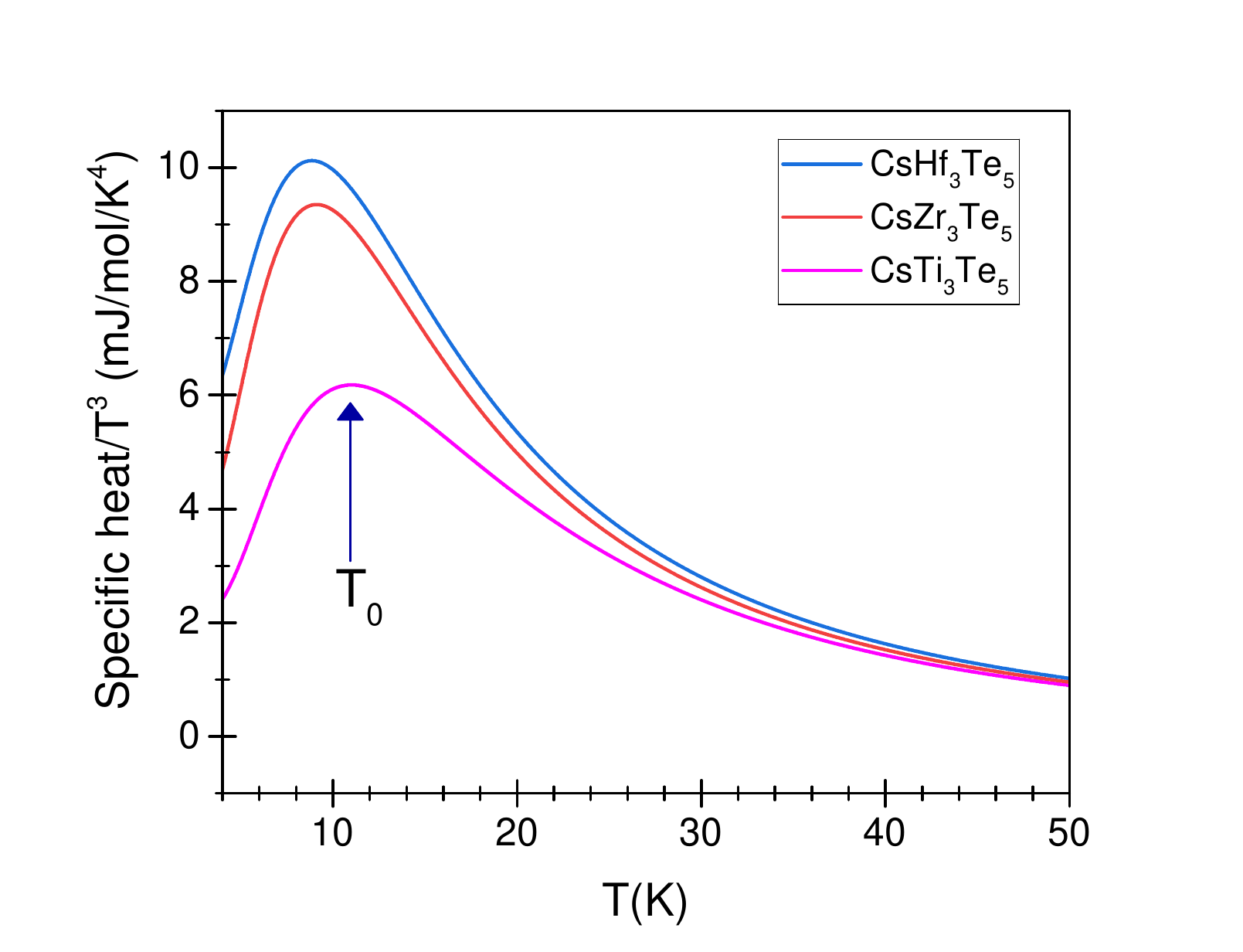}
\caption{Specific heat $C(T)/T^3$ vs.~$T$ profile calculated for the studied kagome metals.}
\label{fig:specific_heat}
\end{figure}

After assessing the elastic, mechanical, and vibrational properties of the studied kagome metals, we turn our attention to the thermodynamic property, specific heat ($C$), and estimate the Debye and Einstein temperatures using the results from our phonon calculations. 
This analysis is particularly important due to the unique dispersion characteristics of the various ``flat" phonon branches.
We note that at low temperatures ($T$), the specific heat at constant pressure ($C_p$) and the specific heat at constant volume ($C_v$) are almost the same, within the experimental uncertainty range~\cite{Cardona2007}. Consequently, in this study, we do not differentiate between $C_p$($T$) and $C_v$($T$).

In the low-$T$ limit, the specific heat $C$($T$) has two primary contributions: a linear term ($\propto T$) from the electronic part and a term proportional to $T^3$ from the crystal lattice. The general relationship is expressed as:
\begin{equation}
    C(T) = \gamma\,T + \beta\,T^3 + \alpha\,T^{-2}.
\end{equation}
Here the last term describes the interaction between the nuclear quadrupole moment and the electric field gradient generated by the electrons and the lattice. 
Usually, this term is negligible even at low $T$. However, it might become significant as $T \rightarrow 0$, particularly below 1\,K~\cite{Serrano2008}.

At low temperatures, $C(T)$ typically follows a $T{^3}$ power law primarily due to the dominant contribution from lattice vibrations. Therefore, plotting $C(T)/T^3$ against $T$ provides a reliable method to discern the role of lattice vibrations on the total heat capacity~\cite{KremerPRB2005, SanchezPRL2007, Cardona2007, Serrano2008, CardonaPRB2009, Singh_elastic}.
The appearance of a peak in the $C(T)/T^3$ vs.~$T$ plot indicates a departure from the Debye behavior, distinguishing the contributions of acoustic and optical phonons to $C(T)$. By noting the position of this peak $T_0$, one can estimate the Einstein temperature $\theta_E$, which is $\theta_E$ $\sim$ 6\,$T_0$~\cite{Cardona2007}. 
The obtained $\theta_E$ values are 66, 54, and 53\,K for CsTi$_3$Te$_5$, CsZr$_3$Te$_5$, and CsHf$_3$Te$_5$, respectively, as listed in Table~\ref{tab:data}. These data suggest that quantum mechanical effects will become dominant in the studied kagome metals below $\theta_E$. 

For comparison, the reported experimental $\theta_E$ ($T_0$) values for bulk Bi and Sb are 45\,K (7.5\,K)~\cite{SanchezPRL2007} and 84\,K (14\,K)~\cite{Serrano2008}, respectively. Additionally, the experimentally reported maximum peak heights of $C(T)/T^3$ at $T_0$ are 2.25 and 0.45 mJ/mol/K$^4$ for bulk Bi and Sb, respectively~\cite{SanchezPRL2007, Serrano2008}. 





Figure~\ref{fig:specific_heat} shows the calculated $C(T)/T^3$ vs.~$T$ profile for the three studied kagome metals. The maximum peak height increases with increasing atomic mass of M atom (Ti $\rightarrow$ Zr $\rightarrow$ Hf). This is due to the fact that Debye temperature $\theta_{Debye}$ decreases with increasing atomic mass of M (see Fig.~\ref{fig:elastic_vel} and Table~\ref{tab:data}) and 
$C \propto (\frac{T}{\theta_{Debye}})^3$.

\subsection{Electronic Properties}

\begin{figure*}[ht]
\centering
\includegraphics[width=2\columnwidth]{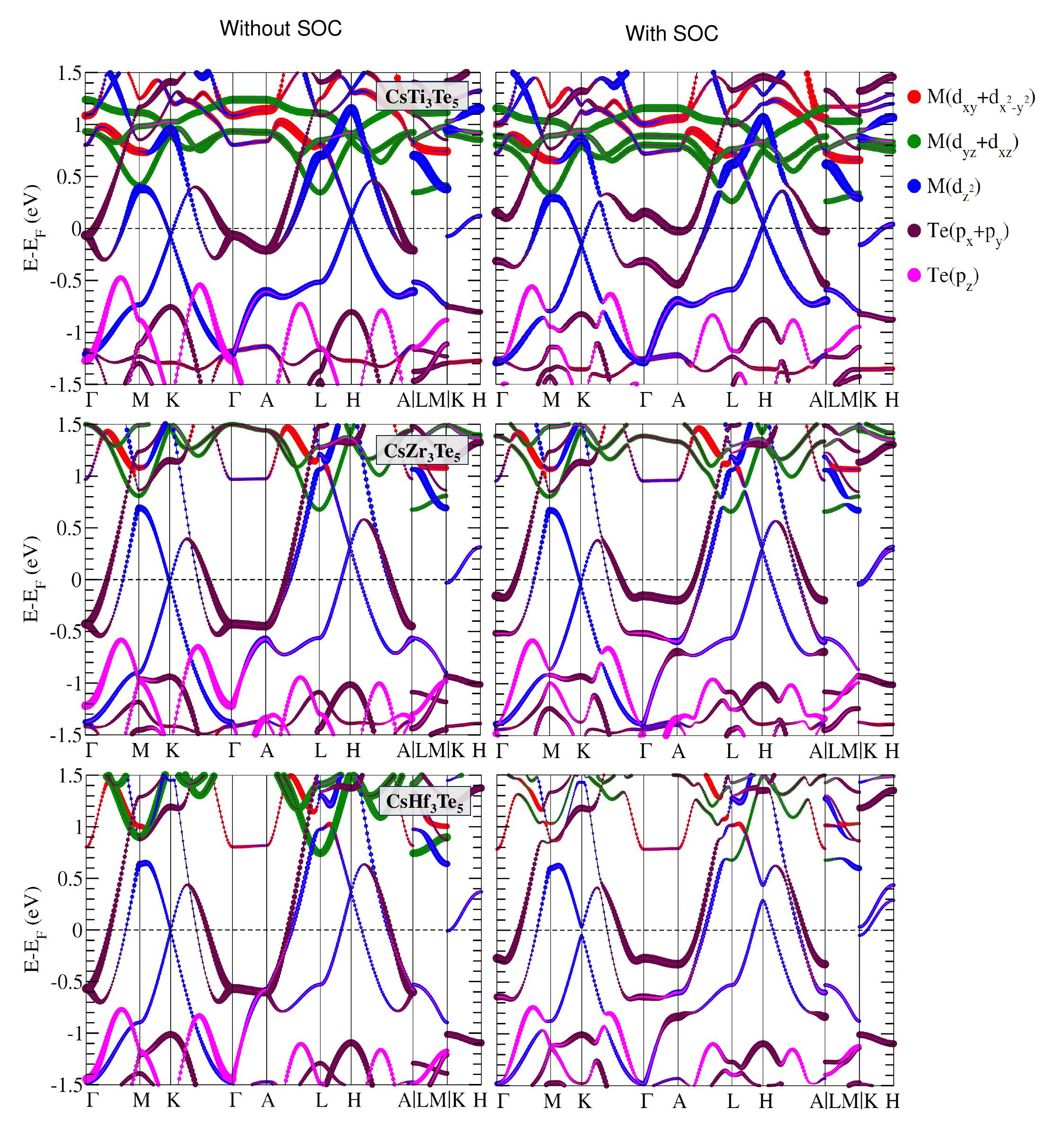}
\caption{Atomic-orbitals projected electronic bandstructure calculated along the high-symmetry direction of the hexagonal Brillouin zone. Left (right) panel shows the bandstructure calculated without (with) including SOC effects.}
\label{fig: Electronic Band Structure with and without SOC }
\end{figure*}

Figure~\ref{fig: Electronic Band Structure with and without SOC } shows the electronic bandstructure of CsTi$_3$Te$_5$, CsZr$_3$Te$_5$, and CsHf$_3$Te$_5$.
The left (right) panels show the bandstructure calculated without (with) SOC, plotted along the high-symmetry directions of the hexagonal Brillouin zone. 
The dotted horizontal line depicts the Fermi level (E$_F$). 
All three investigated systems exhibit metallic behavior with two Dirac-like band crossings at the H and K points of the Brillouin zone, which is typical for kagome metals~\cite{Ye_nature, Ortiz2020, Ghimire_Nature_2020, Yin_nature_2022, Neupert2022}. In CsTi$_3$Te$_5$, the Dirac crossing near the H point is located in the vicinity of the Fermi level. 
However, as the atomic number of the transition metal M increases, this crossing gradually shifts to higher energies above the Fermi level due to the varying electronegativity. 
Further, the inclusion of SOC results in the opening of a band gap at the Dirac crossings. The SOC induced gap is relatively small in CsTi$_3$Te$_5$ and CsZr$_3$Te$_5$, but it becomes significant in CsHf$_3$Te$_5$. This is due to the considerable atomic SOC of the heavier element Hf. 

To further investigate the orbital contributions of different atoms in the electronic band structure near the Fermi level, the orbital-projected electronic band structure of all three compounds is computed and presented in Figure~\ref{fig: Electronic Band Structure with and without SOC }.  
Our results reveal that Te-$p$ and M-$d$ orbitals primarily contribute to the electronic states near the $E_F$, while contributions from the Cs atoms are negligible near $E_F$. 
These findings align well with the recent work of Si~\textit{et al.}~\cite{Si_PRB2022}. 
Even within the Te-$p$ orbitals, the in-plane orbitals $p_{x,y}$ exhibit larger contributions than the out-of-plane $p_z$ orbitals. 
Besides, for the transition-metal atoms, the \( d_{z^2} \) orbitals predominantly contribute near the Fermi level.
Furthermore, Fig.~\ref{fig: Electronic Band Structure with and without SOC } reveals that the in-plane $p_x$ and $p_y$ orbitals of the Te atom are higher in energy than the out-of-plane $p_z$ orbitals due to stronger in-plane steric interactions between atoms, as inferred from the crystal structure of these kagome materials. For the same reason, the in-plane $d$ orbitals of the transition metals are also higher in energy than the out-of-plane $d_{xz}$/$d_{yz}$ orbitals.

\section{Conclusion}
This work provides a comprehensive first-principles investigation of the elastic, mechanical, vibrational, thermodynamical, and electronic properties of kagome metals CsM\(_3\)Te\(_5\) (M = Ti, Zr, or Hf). All three studied materials are dynamically and elastically stable, exhibiting ductile elastic and mechanical properties comparable to the hexagonal systems such as Bi and Sb.   
Additionally, CsTi\(_3\)Te\(_5\) exhibits the highest values of elastic wave velocities (\( v_l \), \( v_t \), and \( v_m \)) and Debye temperature (\( \theta_{\text{Debye}} \)) among the three studied kagome metals. 
%
%
Notably, we observe dispersionless, ``almost flat," phonon branches extending across the whole Brillouin zone, contributing to sharp peaks in the phonon density of states at specific frequencies. 
Among the Raman and IR active phonon modes, the IR active \(\emph{A}\textsubscript{2u}\) and \(\emph{E}\textsubscript{1u}\) phonon modes associated with vibrations of the M atom (Ti, Zr, or Hf) progressively soften with increasing atomic mass of M. 
In contrast, frequencies of other phonon modes remain largely unchanged. 
A detailed analysis of the specific heat data reveals estimated Einstein temperatures of 66, 54, and 53\,K for the studied kagome metals, placing them within the same range as those of hexagonal Bi (45\,K) and Sb (84\,K).
Furthermore, all three studied systems exhibit metallic behavior with characteristic Dirac-like band crossings at the H and K points of the hexagonal Brillouin zone. 
The in-plane \( p_x \) and \( p_y \) (d) orbitals of the Te (M) atom are higher in energy than the out-of-plane \( p_z \) (\( d_{xz}/d_{yz} \)) orbitals , indicating stronger in-plane steric interactions consistent with the crystal structure of these materials. 
Overall, this theoretical investigation sheds light on the intriguing quantum properties of these kagome metals and provides valuable insights for their potential applications in future technologies.

\section*{Acknowledgements}
Authors acknowledge support from the University Research Awards at the University of Rochester. SS is supported by the U.S.~Department of Energy, Office of Science, Office of Fusion Energy Sciences, Quantum Information Science program under Award No.~DE-SC-0020340. 
Authors thank the Pittsburgh Supercomputer Center (Bridges2) supported by the Advanced Cyberinfrastructure Coordination Ecosystem: Services \& Support (ACCESS) program, which is supported by National Science Foundation grants \#2138259, \#2138286, \#2138307, \#2137603, and \#2138296. 

\bibliography{biblio}

\end{document}